\documentclass[aps,pre,twocolumn,nofootinbib,showpacs]{revtex4-1}
\usepackage{latexsym}
\usepackage{amsmath,amsfonts,graphicx,accents}
\usepackage{amsbsy}
\usepackage{hyperref}
\usepackage{mathrsfs}
\usepackage{color}
\usepackage{psfrag}
\usepackage{enumerate}
\usepackage{amsmath,amssymb,calc,amsfonts}
\usepackage{latexsym}
\usepackage[utf8]{inputenc}
\usepackage[percent]{overpic}
\DeclareFontFamily{U}{rsfs}{}         
\DeclareFontShape{U}{rsfs}{m}{n}{<5> rsfs5 <6><7> rsfs7          %
  <8><9><10><10.95><12><14.4><17.28><20.74><24.88> rsfs10}{}     %
\DeclareMathAlphabet{\mathfs}{U}{rsfs}{m}{n}                     %
\definecolor{indiagreen}{rgb}{0.07, 0.53, 0.03}
\def\beq{\begin{eqnarray}}
\def\eeq{\end{eqnarray}}

\def\nn{\nonumber\\}

\def\={\stackrel{\Delta}{=}}

\def\grad{\nabla}

\begin{document}
\title{Black Hole Entropy production and Transport coefficients in Lovelock Gravity}
\author{
  C. Fairoos}\email{fairoos.c@iitgn.ac.in} \affiliation{Indian
  Institute of Technology, Gandhinagar, 382355, Gujarat , India.} \author{
  Avirup Ghosh}\email{avirup.ghosh@iitgn.ac.in} \affiliation{Indian
  Institute of Technology, Gandhinagar, 382355, Gujarat , India.}
\author{ Sudipta Sarkar}
\email{sudiptas@iitgn.ac.in} \affiliation{Indian
  Institute of Technology, Gandhinagar, 382355, Gujarat , India.}
  
   \begin{abstract}
   We study the entropy evolution of black holes in Lovelock gravity by formulating a thermodynamic generalization of null Raychaudhuri equation. We show that the similarity between the expressions of entropy change of the black hole horizon due to perturbation and that of a fluid, which is out of equilibrium, transcends beyond general relativity to the Lovelock class of theories. Exploiting this analogy we find that the shear and bulk viscosities for the black holes in Lovelock theories exactly match with those obtained in the membrane paradigm and also from holographic considerations.

    \end{abstract}

  \maketitle
\section{Introduction}
The realization that black holes follow laws that are similar to the laws of thermodynamics \cite{Bardeen:1973gs} and the discovery that black holes can be endowed with a temperature \cite{Hawking:1974sw} and entropy \cite{Bekenstein:1973ur} opened up an yet unanswered question about the nature of the micro states of a black hole.
It is generally expected that a, still elusive, quantum theory of gravity would account for such micro states. In the absence of any such complete quantum theory of gravity, one resorts to exploring alternative routes that give an effective description of the underlying microscopic dynamics. For example, the membrane paradigm \cite{Price:1986yy,Parikh:1997ma} is an effective description of the black hole physics, from the point of view of an outside observer, in which one models the black hole horizon by a membrane of a fictitious fluid living on the horizon surface. The macroscopic transport coefficients of the fluid are obtained from an intrinsic quasi-local stress-energy tensor associated with the null event horizon and are interpreted as quantities obtained by coarse-graining of `some' quantum theory describing the micro states. This is similar in spirit to the understanding of transport coefficients of ordinary fluids from statistical mechanics. The covariant conservation of the fluid stress-energy tensor gives rise to the energy evolution equations, which when compared to the equations for a relativistic fluid gives the analogous bulk and shear viscosities of the horizon fluid \cite{Price:1986yy,Parikh:1997ma,Zhao:2015inu}.

In the case of fluids, one can proceed further and find the evolution of another macroscopic quantity viz. the entropy current. The entropy evolution equation for physical systems is important in its own right owing to the fact that an increase in entropy naturally defines a thermodynamic arrow of time and brings forth the notion of irreversibility. In the case of fluids, the irreversibility is incorporated in Boltzmann's assumption of molecular chaos i.e the particles constituting the fluid are uncorrelated before a collision. Such a clear notion does not exist for black holes and it is important to find out, to what extent the analogy between the membrane fluids and ordinary fluids hold at the phenomenological level. In other words, does the evolution of black hole entropy has any correspondence to the well known fluid dynamical picture? This is more or less known to hold for general relativity but remains to be shown for other theories of gravity. Though the conclusion is expected to be in the affirmative, it turns out to be actually difficult to establish because of the following reason:

In fluid dynamics, non-equilibrium phenomenon can be studied as a perturbation over a local or global equilibrium state described by an equilibrium density matrix or a distribution function, using either the Chapman and Enskog method or the moment method \cite{Israel:1979wp,groot1980relativistic,cercignani2002relativistic}. At each perturbation order, one then arrives at improved contributions to the evolution equations due to corrections to the transport coefficients. it turns out that, in the Chapman and Enskog scheme,  irreversible effects, determined by the viscosities, arise at first order in the stress-energy tensor of the fluid, while non zero contributions to entropy production arise only at the second order or higher. 

In the case of black holes, the approach taken to study entropy evolution is to perturb an initial stationary black hole and find the evolution of an entropy functional along a perturbed event horizon, in the spirit of the physical process version of the first law \cite{wald1994quantum,Jacobson:2003wv,Amsel:2007mh,Rogatko:2002eu,Bhattacharjee:2014eea}. Interestingly, even in this case, viscous effects do not play a role at linear orders in perturbation. Hence the algebraic complications of second order perturbations in different extensions of general relativity seem inescapable in order to check if the evolution is indeed similar to that of a fluid. Apart from extending the membrane analogy for black holes, it will also point out if the perturbative expansion, in this case, is similar in spirit to the Chapman Enskog scheme adapted to relativistic fluid dynamics. Further, the precise correspondence with the fluid picture will provide us with information regarding the nature of the quantum theory describing the black hole micro-states.\\

General relativity may only make sense as a Wilsonian effective theory with new higher curvature terms in the action. As a result, it is natural to enquire if this fluid-like behavior of a black hole horizon transcends beyond general relativity and valid for any diffeomorphism invariant metric theory of gravity. The first law of black hole mechanics can be extended to all diffeomorphism invariant theories and it is possible to write down an entropy associated with any stationary event horizon. The physical process version of first law which describes the evolution of the entropy of the horizon has been studied for linearized perturbation for various higher curvature theories \cite{Chatterjee:2011wj,Kolekar:2012tq,Bhattacharjee:2015yaa,Bhattacharjee:2015qaa}. Also, the membrane paradigm has been generalized to all Lovelock class of gravity theories \cite{Jacobson:2011dz,Kolekar:2011gg}. But, a major limitation of all these studies is that they are limited to only linear order of the horizon perturbations. To determine the shear and bulk viscosity coefficients of the horizon from the physical process law, we need to consider the terms which are second order in perturbation. In GR, the Raychaudhuri equation for the horizon generators immediately provides us an evolution equation for the Bekenstein-Hawking entropy. The shear and bulk viscosity obtained from the area evolution in GR match exactly with that of the membrane matter. If we want to complete the fluid analogy of black holes beyond general relativity, we need to perform the same calculation for higher curvature theories.  This would require a thermodynamic generalization of the null Raychaudhuri equation. In general relativity, the cross-sectional area of the horizon is proportional to the entropy and we can interpret the null Raychaudhuri equation as an entropy evolution law. Although the Raychaudhuri equation is a geometric result independent of the theory of gravity, this thermodynamic interpretation breaks down in any other metric theory of gravity for which the horizon entropy is no longer proportional to the area. Therefore, we should look for an equation which describes the evolution of the entropy rather than the area. \\

In this work, we complete this calculation and show that the `integrated entropy change' of a dynamical black hole along the perturbed event horizon for all Lovelock class of theories in arbitrary dimensions is similar to that of the entropy evolution for fluids. Moreover, we show that the viscous coefficients obtained from this relation are same as that obtained from the membrane paradigm framework. This demonstrates that the fluid description of black hole horizon transcends beyond general relativity to a well-motivated class of higher curvature theories. \\

\section{General Framework}\label{GENF}
The entropy functional of the black hole horizon can be written in a general form,
\begin{eqnarray}\label{entropy}
S=\frac{1}{4G}\int_{\partial \mathcal H}(1+\rho)\sqrt{h}~d^{D-2}\tau,
\end{eqnarray}
where $\partial\mathcal H$ is a horizon cross-section, $\{ \tau^a\}$ are the coordinates and $h_{ab}$ is the induced metric on $\partial\mathcal H$. For general relativity, we set $\rho=0$.

The general framework involves taking the variation of this function along the generators $k$ of the horizon $\mathcal H$. Note that in arbitrary dimensions the cross sections of the horizon may not be closed manifolds e.g black brane like solutions. In such cases, we will assume that the integration in $\tau's$ is over an open-ball in $\mathbb R^{D-2}$, so that we can throw away certain total divergence terms. One can choose the generators to be affinely parametrized by the parameter $\lambda$ (say). The change of entropy between two cross-sections of the horizon is then given as,
\begin{eqnarray}
\Delta S=\int_{i}^{f}\Theta \, \sqrt{h}\, d\lambda \,d^{D-2}\tau ,
\end{eqnarray}
where, $\Theta$ represents the change of entropy per unit cross sectional area and is defined as,
\begin{eqnarray}
\Theta =\frac{1}{4 G} \left( \frac{ d \rho}{d \lambda} +  \theta^{(k)} + \rho \,\theta^{(k)} \right)
\end{eqnarray}
In GR, $\Theta = \frac{\theta^{(k)}}{4 G}$ where $\theta^{(k)}$ is the expansion of the generators of the event horizon. For more general theories, there is no such geometric interpretation. Doing an integration by parts, one arrives at an expression for the change in entropy in terms of the second derivative of the entropy functional,
\begin{eqnarray}
\Delta S=\left[\lambda \Theta \right]_{i}^{f}-\int_{i}^{f} \lambda \left(\frac{d\Theta}{d\lambda} + \theta^{(k) }\Theta \right)\, \sqrt{h}\, d\lambda \,~d^{D-2}\tau
\end{eqnarray}
Note that if one integrates from the initial bifurcation surface, where $\lambda$ can be chosen to be zero, to a final stationary black hole, where $\Theta$ is zero, the first term does not contribute. This is the choice of initial and final cross-sections we will be making throughout this paper. However, for arbitrary choices of cross-sections, there is an interpretation of this term in terms of the energy of the membrane fluid \cite{Chakraborty:2017kob}. The final expression for the entropy change between an initial bifurcation surface to final stationary cross-section is,

\begin{eqnarray}
\Delta S=-\int_{i}^{f} \lambda \left(\frac{d\Theta}{d\lambda} + \theta^{(k)} \Theta \right)\, \sqrt{h}\, d\lambda \,~d^{D-2}\tau\label{entropy_change}
\end{eqnarray}
In the next sections, we will use this equation and evaluate the entropy change for various Lovelock class of theories and try to make a connection with the entropy evolution of a fluid.
\\
\section{Einstein-Gauss Bonnet gravity}

Let us consider the simplest generalization of general relativity: The Einstein Gauss Bonnet gravity where the action is given as,
\beq
\mathscr S&=&\frac{1}{16\pi G}\int\sqrt{-g}R+\nn
&&~\sqrt{-g}\alpha\underbrace{\big(R^2-4R^{\mu\nu}R_{\mu \nu}+R^{\mu\nu\rho\sigma}R_{\mu\nu\rho\sigma}}_{\mathcal L_{GB}}\bigg)d^Dx
\eeq
where $\alpha$ is a constant. The equations of motion that follow from the above action are,
\beq
G_{\mu \nu}=\alpha H_{\mu\nu}+8\pi G T_{\mu\nu},
\eeq
where $H_{\mu\nu}=-2\big(RR_{\mu\nu}-2R_{\mu}~^{\rho}R_{\rho\nu}-2R^{\rho\sigma}R_{\mu\rho\nu\sigma}+R_{\mu}~^{\rho\sigma\alpha}R_{\nu\rho\sigma\alpha}\big)+\frac{1}{2}g_{\mu\nu}\mathcal L_{GB}$. The entropy of a dynamical black hole in Gauss-Bonnet gravity is obtained by fixing the Jacobson-Kang-Myers (JKM) ambiguities \cite{Jacobson:1995uq} in Wald entropy \cite{Wald:1993nt,Iyer:1994ys}, by demanding that a local entropy increase law holds at linear order in perturbation from an initial stationary state and is given by \cite{Bhattacharjee:2015yaa},

\begin{eqnarray}
S=\frac{1}{4G}\int_{\partial \mathcal H}(1+2\alpha\mathcal R)\sqrt{h}~d^{(D-2)}\tau,
\end{eqnarray}
where $\mathcal R$ is the intrinsic Ricci scalar of the horizon cross-sections. Here, we will be interested in finding the full all order evolution of it along the generators of the dynamical event horizon. 
The complete expression for $\frac{d\Theta}{d\lambda}$ is given by,
\begin{widetext}
\begin{equation}\label{fullexp}
\begin{gathered}
4 G \frac{d\Theta}{d\lambda}=-\frac{\theta^{(k)2}}{D-2}-\sigma^{(k)ab}\sigma^{(k)}_{ab}-6\alpha\frac{(D-4)\theta^{(k)2}\mathcal R}{(D-2)^2}-2\alpha\sigma^{(k)ab}\sigma^{(k)}_{ab}\mathcal R-4\alpha\frac{(D-8)\theta^{(k)}\sigma^{(k)ab}\mathcal R_{ab}}{(D-2)}\\
+8\alpha\sigma^{(k)a}_c\sigma^{(k)cb}\mathcal R_{ab}-4\alpha \mathcal R_{fabp}~\sigma^{(k)ab}\sigma^{(k)pf}\\
+2\alpha\Bigg[2\bigg(D_c\beta^c\bigg)\bigg(K^{(k)}_{ab}K^{(k)ab}\bigg)-4\bigg(D_c\beta^b\bigg)\bigg(K^{(k)}_{ab}K^{(k)ac}\bigg)+2\beta^c\beta_cK^{(k)}_{ab}K^{(k)ab}-4\beta_cK^{(k)}_{ab}\beta^bK^{(k)ac}\Bigg]\\
+4\alpha \bigg[2\bigg(D^b\beta^f\bigg)\bigg(K^{(k)}K^{(k)}_{bf}\bigg)-2\bigg(D_a\beta^a\bigg)\bigg(K^{(k)}\bigg)^2+2h^{ab}\beta^cK^{(k)}_{ac}\beta_bK^{(k)}-h^{ab}\beta_a\beta_b(K^{(k)})^2\bigg]\\
+4\alpha R_{kk}\frac{(D-3)(D-4)\theta^{(n)}\theta^{(k)}}{(D-2)^2}-4\alpha h^{ac}h^{bd}R_{kckd}\frac{(D-4)\theta^{(k)}\sigma^{(n)}_{ab}}{D-2}-4\alpha h^{ac}h^{bd}R_{kckd}\frac{(D-4)\theta^{(n)}\sigma^{(k)}_{ab}}{D-2}\\
+8\alpha h^{ac}h^{bd}R_{kckd}\sigma^{(k)}_{af}\sigma^{(n)f}_b-4\alpha R_{kk}\sigma^{(k)}_{ab}\sigma^{(n)ab}-8\pi G\, T_{kk}
+\alpha\, \text{(total derivatives)}
\end{gathered}
\end{equation}
\begin{equation*}
\begin{gathered}
\end{gathered}
\end{equation*}
\end{widetext}

To comprehend this formidable equation, let us first spell out the notations. The horizon is generated by the null vector $k^\mu$ and the space-time metric has been decomposed as $ g_{\mu\nu} = h_{\mu\nu} -k_\mu n_\nu - k_\nu  n_\mu$ where $n^\mu$ is an auxiliary null vector. Together $k^\mu$ and $n^\mu$ are the two null normals of the horizon cross-section. $\theta^{(k)}$ and $\sigma^{(k)}_{ab}$ are the expansion and shear of the horizon congruence respectively.  $\theta^{(n)}$ and $\sigma^{(n)}_{ab}$ are the same for the null congruence generated by $n^\mu$. $K^{(i)}_{ab}$ is the extrinsic curvature of the horizon cross section w.r.t the null normal $i = k , n$. We have also used the notations $R_{k c  k d} = R_{\mu c  \rho d } k^\mu k^\rho$, $ \beta_a = - n^\mu \grad_a k_\mu$ etc. Setting $ \alpha= 0$, we will obtain the familiar null Raychaudhuri equation. Otherwise, this equation is the thermodynamics generalization of the null Raychaudhuri equation \footnote{For a different approach to study Raychaudhuri equation in higher curvature gravity, see \cite{Burger:2018hpz}.}. The expansion and shear of the horizon generators i.e. $\theta^{(k)}$ and $\sigma^{(k)}_{ab}$ vanish on the background stationary horizon and therefore are at least linear order in perturbation. But, the expansion of the auxiliary null vector $\theta^{(n)}$ is non zero even on the stationary horizon. The total derivative terms involve spacial derivative of the extrinsic curvatures and are second order in perturbation. If we consider only the terms linear in perturbation, we would obtain:
\begin{equation}
\frac{ d \Theta}{d \lambda} = - 2 \pi \, T_{kk} + {\cal O}(\epsilon^2)
\end{equation}
This is the equation which will give us the linearized version of the physical process law which ensures that if the null energy condition $(T_{\mu\nu}k^\mu k^\nu \geq 0)$ holds, the entropy always increases \cite{Chatterjee:2011wj, Kolekar:2012tq, Bhattacharjee:2015yaa, Bhattacharjee:2015qaa}. It is intriguing that the linearized version of the entropy evolution equation for Einstein-Gauss-Bonnet gravity is identical in form with the linearized Raychaudhuri equation even though the entropy density and the equation of motion now contain non-trivial corrections.  \\

The full equation will give an exact expression of the change of horizon entropy. We would like to apply this equation to understand the full evolution of horizon entropy. Due to the complex structure of the terms, it is difficult to obtain any conclusion in general. So, to make sense of this equation, we will now specialize in the case of general second order perturbations about a static black hole background with maximally symmetric horizon cross-section. This choice fulfills our purpose to separate out the dissipative bulk and shear viscous coefficients. Note that, we keep the perturbation completely arbitrary. We will consider changes of the entropy from the initial bifurcation surface to the final stationary cross-section. Then, in this \textit {order by order} in  $\theta^{(k)}$ and $\sigma^{(k)}_{ab}$ calculation, we will obtain followings up to second order: 

\begin{widetext}
\beq\label{entropychange}
\frac{\kappa_H}{2\pi}\Delta^{(1)}\mathcal S&=&\int_{i}^{f}\accentset{1}{T}_{\mu\nu}\xi^\nu d\Sigma^\mu\\
\frac{\kappa_H}{2\pi}\Delta^{(2)}\mathcal S&=&\frac{1}{16\pi G}\int_{i}^{f}\bigg[-2\theta^{(k)2}\bigg(\frac{D-3}{D-2}\bigg)\bigg(1-4\alpha(D-4)\frac{\kappa_H}{r_H}+\frac{2\alpha\accentset{0}{\mathcal R}(D-4)(D-5)}{(D-2)(D-3)}\bigg)\nn
&&~~~~~~~~~~~~~~~~~~+2\sigma^{(k)2}\bigg(1-4\alpha(D-4)\frac{\kappa_H}{r_H}+\frac{2\alpha\accentset{0}{\mathcal R} (D-4)(D-5)}{(D-2)(D-3)}\bigg)\bigg]\sqrt{h}d^{(D-2)}\tau~dt \nn
&&~~~~~~~~~~~~~~~~~~~~
+\int_{i}^{f}\accentset{1}{T}_{\mu\nu}\widetilde{\xi^\nu}d\Sigma^\mu + \int_{i}^{f}\accentset{2}{T}_{\mu\nu}\xi^\nu d\Sigma^\mu,
\eeq
\end{widetext}
The superscripts in the expression refer to the order of perturbation and $\kappa_H$, $r_H$ are the surface gravity and radius of the background horizon, respectively. For maximally symmetric background one has $\accentset{0}{\mathcal R}_{abcd}=\frac{\accentset{0}{\mathcal R}}{(D-2)(D-3)}(\accentset{0}{h}_{ac}\accentset{0}{h}_{bd}-\accentset{0}{h}_{bc}\accentset{0}{h}_{ad})$. Moreover, on the background the $\beta_a's$ can be set to zero. To understand the ordering, let us assume that the background Killing vector $\xi^\mu$ ($\partial_{t}$) is fixed while the null generators of the perturbed horizon get modified at each order such that,
\begin{eqnarray}
k=\accentset{0}{k}+\accentset{1}{k}+...,
\end{eqnarray}
where
\begin{eqnarray}
\accentset{0}{k}=\accentset{0}{f}\xi+\accentset{0}{V},~~~~~\accentset{1}{k}=\accentset{1}{f}\xi+\accentset{1}{V},
\end{eqnarray}
and so on. $V's$ are vectors tangent to the cross-sections of the horizon. The functions $f's$ and vectors $V's$ are given as $\accentset{(0)}{f}=1/\kappa\accentset{0}{\lambda},~~\accentset{0}{V}=0$, while $\accentset{1}{f},~\accentset{(1)}{V}$ are determined by the perturbation. The parameter $t$ along $\xi$, to zeroth order, is $t =\frac{\log\accentset{0}{\lambda}}{\kappa_H}$. For a maximally symmetric static background we also have $\theta^{(n)}=-\frac{(D-2)\kappa_H}{r_H}\accentset{0}{\lambda}$. Writing $R_{kk},~R_{kckd}$ in terms of derivatives of the extrinsic curvature, one arrives at the $\kappa_H$ dependent terms. Further, $T_{\mu\nu}$ in general have all orders. Hence we can expand it as,
\begin{eqnarray}
T_{\mu\nu}=\accentset{0}{T}_{\mu\nu}+\accentset{1}{T}_{\mu\nu}+\accentset{2}{T}_{\mu\nu}+...,
\end{eqnarray}
where $\accentset{(0)}{T}_{\mu\nu}$ is the background stress energy tensor which satisfies the condition $ \accentset{(0)}{T}_{kk} = 0$ so that there is no flux though the background stationary horizon. Using this expression, one can arrange the $T_{\mu\nu}$ terms as in (eq.\ref{entropychange}). Note that $\tilde\xi$ is a combination of $f's$, $k's$ and $V's$ and  $\theta^{(k)}$ and $\sigma^{(k)}_{ab}$ are now the expansion and shear for the non-affinely parametrized null geodesics ($\xi$). \\

The first expression is the usual first-order expression for the physical process version of the first law. If we identify $ \frac{\kappa_H}{2\pi}$ as the Hawking temperature ($T_H$) of the background stationary horizon, this reduces to $T_H \Delta^{(1)}\mathcal S = \Delta^{(1)} Q$. This is the Clausius relationship which shows that in the linear order, the entire change of entropy is due to the external heat flux. There is no dissipation or entropy production in this order and the entire first order heat flux is balanced by the change the entropy of the horizon.\\

The more interesting one is the second equation. The terms in this equation represent viscous effects namely the entropy production due to the bulk and shear viscosities. If one compares the contribution to this equation coming from viscous flow,
\beq
\frac{\kappa_H}{2\pi}\Delta^{(2)}\mathcal S&=&\int_{i}^{f}\bigg[\zeta\theta^{(k)2}+2\eta\sigma^{(k)2}\bigg]\sqrt{h}d^{D-2}\tau~d t\nn \nonumber \\ &+& \Delta Q, \label{2ndorder}
\eeq
then one can read off the bulk and the shear viscosities as,
\beq
\eta &=&\frac{1}{16\pi G}\bigg(1-4\alpha(D-4)\frac{\kappa_H}{r_H}+\frac{2\alpha\accentset{0}{\mathcal R} (D-4)(D-5)}{(D-3)(D-2)}\bigg),\nn
\zeta &=&-2\bigg(\frac{D-3}{D-2}\bigg)\eta.
\eeq
The general relativity limit can be easily obtained by setting $\alpha = 0$. So the thermodynamic generalization of the Raychaudhuri equation for Einstein gauss Bonnet gravity gives us exactly the same structure of the entropy change as in general relativity in the second order with a corrected expression of shear and bulk viscosities. The viscosity contributes to the entropy production only at the second order of perturbation. Note that, these were previously calculated in the framework of membrane paradigm \cite{Jacobson:2011dz}, and remarkably our result exactly matches with the expression in \cite{Jacobson:2011dz}.

If the background is a $D$ dimensional black brane solution with cosmological constant $ \Lambda = - (D -1) (D-2) / L^2$, the ratio of this shear viscosity and entropy density $s$  would be 
\beq
 \frac{\eta}{ s} = \frac{1 }{ 4 \pi} \left( 1 - \frac{ 2  \alpha ( D - 1)(D - 4)}{L^2} \right).
 \eeq
This result has been obtained previously in \cite{Brigante:2007nu} in the AdS/CFT context, for the dual fluid residing on the AdS boundary. As is evident from the above expression,  the Kovtun-Starinets-Son (KSS) bound,  $\frac{\eta}{ s} \geq \frac{1 }{ 4 \pi}$, \cite{Kovtun:2004de} is violated for $\alpha >0$. Causality constraints in the bulk and energy positivity in the dual CFT do impose constraints on the allowed range of Gauss Bonnet coupling $\alpha$ \cite{Camanho:2009vw}, but there is still a window  for the values of $\alpha$ for which the KSS bound is violated although all the other constraints are satisfied.  The study of perturbation of the fluid on the AdS boundary of a black brane solution in Einstein-Gauss-Bonnet Gravity \cite{Buchel:2009sk} also leads to the same expression. Interestingly, the dynamics of the horizon fluid produces the same expression for the shear viscosity to entropy density ratio.

\section{Lovelock Gravity}
Let us now discuss the extension of this result to the full Lovelock gravity. The Lovelock class of Lagrangians are the unique extension of the Einstein Hilbert action, involving higher curvature interactions, that yields equations of motion which are of second degree in derivatives of $g_{\mu\nu}$. The horizon entropy density obtained after fixing the JKM ambiguities (using the linearized second law) is given by (eq. \ref{entropy}) with $\rho$ given by \cite{Sarkar:2013swa},
\begin{equation}
\rho=\bigg(\sum_{m=2}^{\frac{D-1}{2}}16\pi Gm\alpha_m~^{D-2}\mathcal L_{m-1}\bigg),
\end{equation}
where $^{D-2}\mathcal L_{m-1}$ is the $(m-1)-th$ order Lovelock Lagrangian evaluated with the intrinsic Riemann tensor $\mathcal R_{abcd}$. In the above expression the Einstein-Hilbert part corresponds to $ m = 1$ and is not included in the sum. 
\begin{widetext}
The second variation of the above entropy has the following form,
\begin{eqnarray}\label{2ndvarL}
\begin{gathered} 
\delta_k^2 S=\frac{1}{4G}\int d^{D-2}x\bigg[\sqrt{h}\bigg(-\frac{\theta^{(k)2}}{D-2}-\sigma^2-R_{kk}\bigg)\bigg(1+\sum_{m=2}^{\frac{D-1}{2}}16\pi Gm\alpha_m~^{D-2}\mathcal L_{m-1}\bigg)\\
+2\,\theta^{(k)}\sqrt{h}\bigg(\sum_{m=2}^{\frac{D-1}{2}}32\pi Gm\alpha_m~^{D-2}\mathscr R^{ab}_{m-1}K^{(k)}_{ab}\bigg)+\theta^{(k)2}\sqrt{h}\bigg(1+\sum_{m=2}^{\frac{D-1}{2}}16\pi Gm\alpha_m~^{D-2}\mathcal L_{m-1}\bigg)\\
+\sqrt{h}\bigg(\sum_{m=2}^{\frac{D-1}{2}}32\pi Gm\alpha_m \delta_k\big( ~^{D-2}\mathscr R^{ab}_{m-1}\big)K^{(k)}_{ab}\bigg)+\sqrt{h}\bigg(\sum_{m=2}^{\frac{D-1}{2}}32\pi Gm\alpha_m~^{D-2}\mathscr R^{ab}_{m-1}\bigg)\bigg(-K^{(k)}_{ac}K^{(k)c}_b+R_{kakb}\bigg)\bigg],
\end{gathered}
\end{eqnarray}
where the notation $~^{D-2}\mathscr R_{ab}$ in the above expression has been introduced for convenience and is given as,
\beq
~^{D-2}\mathscr R^a_{(m-1)~b}&=&\frac{1}{2\times 16\pi G}\frac{m-1}{2^{m-1}}\delta^{a_1b_1a_2b_2....a_{m-1}b_{m-1}}_{bd_1c_2d_2....c_{m-1}d_{m-1}}\mathcal R^{ad_1}_{a_1b_1}\mathcal R^{c_2d_2}_{a_2b_2}...\mathcal R^{a_{m-1}b_{m-1}}_{c_{m-1}d_{m-1}}\nn
&=&^{D-2}E^a_{(m-1)~b}+\frac{1}{2}\delta^a~_{b}~^{D-2}\mathcal L_{m-1},
\eeq 
where $^{D-2}E^a_{(m-1)~b}$ is the $(m-1)$ th order Lovelock equations of motion in $(D-2)$ dimensions.
\end{widetext}
There are large number of algebraic steps that one has to go through. We merely state the expressions used and steps, here.  The $R_{kk}$ term in eq. \eqref{2ndvarL} is replaced using the equations of motion. On using the Gauss' equation \eqref{gauss} the expression for $R_{kk}$ reads,
\begin{widetext}
\begin{eqnarray}\label{EOM}
\begin{gathered}
-R_{kk}+8\pi G~T_{kk}=\sum_{m=2}^{\frac{D-1}{2}}-\frac{2m\alpha_m}{2^{m}}\delta^{ba_1b_1...a_{m-1}b_{m-1}}_{dc_1d_1...c_{m-1}d_{m-1}}R^{nd}_{kb}\sum_{k=0}^{m-1}~^{m-1}C_k\underbrace{(KK)^{c_1d_1}_{a_1b_1}...(KK)^{c_{k}d_{k}}_{a_{k}b_{k}}}_{k-terms}\mathcal R^{c_{k+1}d_{k+1}}_{a_{k+1}b_{k+1}}...\mathcal R^{c_{m-1}d_{m-1}}_{a_{m-1}b_{m-1}}\\
+\sum_{m=2}^{\frac{D-1}{2}}\frac{2m(m-1)\alpha_m}{2^{m}}\delta^{bcfa_1b_1...a_{m-2}b_{m-2}}_{adec_1d_1...c_{m-2}d_{m-2}}R^{na}_{bc}R^{de}_{kf}
\times\sum_{k=0}^{m-2}~^{m-2}C_k\underbrace{(KK)^{c_1d_1}_{a_1b_1}...(KK)^{c_{k}d_{k}}_{a_{k}b_{k}}}_{k-terms}\mathcal R^{c_{k+1}d_{k+1}}_{a_{k+1}b_{k+1}}...\mathcal R^{c_{m-2}d_{m-2}}_{a_{m-2}b_{m-2}},
\end{gathered}
\end{eqnarray}
where $(KK)^{cd}_{ab}=h^{ce}h^{df}\big(K^{\mu}_{ae}K_{\mu bf}-K^{\mu}_{af}K_{\mu be}\big)$ is a product of extrinsic curvatures as it appears in Gauss' equation eq. \eqref{gauss}. Also, it should be noted that,
\begin{eqnarray}
&R_{kakb}\bigg(-\frac{1}{2}h^{ab}\sum_{m=2}^{\frac{D-1}{2}}32\pi Gm\alpha_m~^{D-2}L_{m-1}+\sum_{m=2}^{\frac{D-1}{2}}32\pi Gm\alpha_m~^{D-2}\mathscr R^{ab}_{m-1}\bigg)\nn
&~~~~~~~~~~~~~~~~~=R_{kakc}h^{bc}\sum_{m=2}^{\frac{D-1}{2}}\frac{2m\alpha_m}{2^m}\delta^{aa_1b_1...a_{m-1}b_{m-1}}_{bc_1d_1...c_{m-1}d_{m-1}} \mathcal R^{c_1d_1}_{a_1b_1}...\mathcal R^{c_{m-1}d_{m-1}}_{a_{m-1}b_{m-1}}
\end{eqnarray}

Further the $R^{na}_{bc}R^{de}_{kf}$ in eq. \eqref{EOM} is replaced using Codazzi's equation eq. \eqref{codazzi}. Finally, the $K_{ab}\delta_k\big(~^{D-2}\mathscr R^{ab}_{m-1}\big)$ term has the following form.
\begin{gather}
\sum_{m=2}^{\frac{D-1}{2}}32\pi Gm\alpha_m 2K^{(k)}_{ab}K^{(k)bc} ~^{D-2}\mathscr R^a_{(m-1)c}+K^{(k)}\bigg(\sum_{m=2}^{\frac{D-1}{2}}32\pi Gm\alpha_m~^{D-2}\mathscr R^{ab}_{m-1}K^{(k)}_{ab}\bigg)\nn
+\sum_{m=2}^{\frac{D-1}{2}}4 m\alpha_m K^{(k)b}_{a}\frac{(m-1)}{2^m}\delta^{aa_1b_1a_2b_2...a_{m-1}b_{m-1}}_{bc_1d_1c_2d_2...c_{m-1}d_{m-1}} h^{d_1f}h^{c_1p} D_{a_1}\Big(D_fK^{(k)}_{b_1p}-D_pK^{(k)}_{b_1f}\Big)\times\mathcal R_{a_2b_2}^{c_2d_2}...\mathcal R_{a_{m-1}b_{m-1}}^{c_{m-1}d_{m-1}}\notag\\
+\sum_{m=2}^{\frac{D-1}{2}}4 m\alpha_m K^{(k)b}_{a}\frac{(m-1)}{2^m}\delta^{aa_1b_1a_2b_2...a_{m-1}b_{m-1}}_{bc_1d_1c_2d_2...c_{m-1}d_{m-1}} K^{(k)d_1f}\mathcal R^{c_1}~_{fa_1b_1}\mathcal R_{a_2b_2}^{c_2d_2}...\mathcal R_{a_{m-1}b_{m-1}}^{c_{m-1}d_{m-1}}
\end{gather}
\end{widetext}
Using these one can get an expression for $\delta_k^2 S$ to complete generality, as in the case of Gauss-Bonnet gravity. The results will not be quoted here, rather we will directly take the second order perturbation approximation about a maximally symmetric stationary Black hole solution. Therefore terms which are at most quadratic in $\theta^{(k)}$ and $\sigma_{ab}^{(k)}$ are retained and the rest are substituted by values they take on the background, by using eq. \eqref{MAX}. Finally $R^{nd}_{kb}$ is replaced using eq. \eqref{EXCE} and a total $\lambda$ derivative extracted.
Following the steps outlined in sec. \ref{GENF} one then gets an analogous expression for entropy change upto second order, 
\begin{eqnarray}\label{ECL}
\Delta S&=&\bigg[V\sqrt h\bigg]_i^f+2\pi \int_{i}^{f}\bigg[\zeta\theta^{(k)2}+2\eta\sigma^{(k)2}+T_{kk}\bigg]\nn
&&~~~~~~~~~~~~~~~~~~~~~~~~~~~~~~~\times\sqrt{h}~d^{D-2}\tau~\lambda d \lambda,
\end{eqnarray}

where $V$ is some function of the geometric quantities at hand and $\eta$, $\zeta$ are given by,
\begin{eqnarray}
\eta&=&\sum_{m=1}^{\frac{D-1}{2}}\alpha_m m\bigg(\frac{1}{D-2}+\frac{2(m-1)}{(D-2)(D-3)}\nn
&&~~~~~~~~~~~~~-\frac{2\kappa_H}{r_H\accentset{0}{\mathcal R}}(m-1)\bigg)(D-2m)~^{D-2}\accentset{0}{\mathcal L}_{m-1},\nn
\zeta&=&-2\bigg(\frac{D-3}{D-2}\bigg)\eta,
\end{eqnarray}
 As is the case with Gauss Bonnet gravity, the recovery of the entropy production equation eq. \eqref{entropychange}, now, crucially depends on the fact that $V$ is zero both at the final stationary state and the initial bifurcation surface. As is evident from eq. \eqref{ECL}, the transport coefficients thus obtained again match with those obtained in the membrane paradigm picture \cite{Kolekar:2011gg}.\\
\section{Discussion}
In general relativity, it is well known that the bulk viscosity coefficient for black hole horizon is negative. This is a consequence of the teleological nature of the event horizon and the use of future boundary condition. Note that, even in standard thermodynamics, there is a close connection between the assumption of causality and the second law of thermodynamics. The requirement of the validity of the second law of thermodynamics with future ``acausal" boundary condition requires opposite sign for the viscosity coefficient \cite{PhysRevE.53.5808}. Our result shows that this interpretation holds for the entire Lovelock class of theories and the bulk viscosity is always of opposite sign of shear viscosity coefficient. In fact, the ratio of the bulk to the shear viscosity coefficients is identical for all Lovelock orders. Mathematically, this is due to the fact that for perturbations of a maximally symmetric horizon, the change of the generalized expansion $\Theta$ is proportional to the change of the geometric expansion. Since the coefficient of the dissipation terms in original Raychaudhuri equation bears the same ratio, we are also obtaining the identical ratio for the bulk and shear viscosities \footnote{ We thank Ted Jacobson for this observation.} . It would be nice if the universality of this ratio can be understood from the microscopic point of view. \\

We would also like to mention the result of \cite{Camanho:2014apa}, where it is shown that for any value of the higher curvature coupling, the causality can be violated by constructing a closed time like curve involving two shock waves. At this stage, the implication of that result for entropy evolution is not very clear. 

Eq.( \ref{fullexp}) contains the full expression of the change of entropy density of the horizon due to the perturbations. It is important to note that this equation is obtained without any assumption on the nature of the perturbations. In principle, this equation can be used to find out whether a generalization of the Hawking area theorem can be established for the the Lovelock class of theories. This may be done by using a future boundary condition $\Theta \to 0$ in the asymptotic future and then by establishing that $d\Theta / d\lambda  < 0$ for matter satisfying an appropriate energy condition. We already know that the increase of entropy for Einstein gauss Bonnet theory requires certain bound on the coupling constant even when we consider only spherically symmetric perturbation \cite{Bhattacharjee:2015qaa}. Also, for generic case, the entropy evolution equation will contain terms which would represent the coupling of viscous dissipations with the background curvature. In fact, there could be terms which are higher order in perturbation. So, to fully understand the entropy evolution equation, we would require many more transport coefficients. This is evident from the expressions obtained here prior to simplification made by assuming a maximally symmetric black hole background. Therefore the dynamics of black hole horizons in higher curvature gravity seems to be similar to higher order hydrodynamics. Obviously, a generalization of our results to other higher curvature theories will provide a better understanding about this intriguing relationship. 

\section{Acknowledgement}
We thank Srijit Bhattacharjee, Arpan Bhattacharyya and Aninda Sinha for extensive discussion. We also thank Jose Edelstein for discussion. AG is supported by SERB, Government of India
through the NPDF grant (PDF/2017/000533). Research of SS is supported by the SERB, Government of India under the SERB MATRICS Grant (MTR/2017/000399). CF thanks University Grant commission for JRF Fellowship.

\appendix
\section{Identities used in the calculations}
\subsection*{Evolution equation for Extrinsic curvature}
For affinely parametrized $k$ i,e $\nabla_kk=0$,
\beq\label{EXCE}
\delta_k K^{(k)}_{ab}=\frac{d}{d\lambda} K^{(k)}_{ab}=-K^{(k)}_{ac}K^{(k)c}_{b}+R_{kakb}
\eeq

\subsection*{Equation of Gauss}
The Gauss Eq. is given by,
\beq\label{gauss}
R_{abcd}=\mathcal R_{abcd}+K^{\mu}_{ac}K_{\mu bd}-K^{\mu}_{ad}K_{\mu bc}
\eeq
\\
\subsection*{Equation of Codazzi}
The Codazzi Eq. is,
\beq\label{codazzi}
R_{kabc}=D_bK^{(k)}_{ac} - D_cK^{(k)}_{ab}+\beta_cK^{(k)}_{ab}-\beta_bK^{(k)}_{ac},
\eeq
where $\beta_a=-n^\mu\nabla_a k_\mu$.\\
\\
\subsection*{Identities for maximally symmetric background}
We have used the followings:
\beq\label{MAX}
\begin{gathered}
~^{D-2}E^{ab}_{m-1}=-\frac{h^{ab}}{2}\frac{D-2m}{D-2} ~^{D-2}\mathcal L_{m-1},\\
\\
\delta^{aa_1b_1a_2b_2......a_{m-1}b_{m-1}}_{bc_1d_1c_2d_2......c_{m-1}d_{m-1}}\mathcal R_{a_2b_2}^{c_2d_2}......\mathcal R_{a_{m-1}b_{m-1}}^{c_{m-1}d_{m-1}}\notag\\
~~~~~~~~~~~~~~~~~=\frac{16\pi G 2^{m-2}(D-2m)^{D-2}\mathcal L_{m-1}}{(D-2)\mathcal R}\delta^{aa_1b_1}_{bc_1d_1}.
\end{gathered}
\\
\end{eqnarray}


\begin{thebibliography}{99}
\bibitem{Bardeen:1973gs} 
  J.~M.~Bardeen, B.~Carter and S.~W.~Hawking,
  ``The Four laws of black hole mechanics,''
  Commun.\ Math.\ Phys.\  {\bf 31}, 161 (1973).
  doi:10.1007/BF01645742
  
\bibitem{Hawking:1974sw} 
  S.~W.~Hawking,
  ``Particle Creation by black holes,''
  Commun.\ Math.\ Phys.\  {\bf 43}, 199 (1975)
  Erratum: [Commun.\ Math.\ Phys.\  {\bf 46}, 206 (1976)].
  doi:10.1007/BF02345020
  
\bibitem{Bekenstein:1973ur} 
  J.~D.~Bekenstein,
  ``Black holes and entropy,''
  Phys.\ Rev.\ D {\bf 7}, 2333 (1973).
  doi:10.1103/PhysRevD.7.2333
  
\bibitem{Price:1986yy} 
  R.~H.~Price and K.~S.~Thorne,
  ``Membrane Viewpoint on Black Holes: Properties and Evolution of the Stretched Horizon,''
  Phys.\ Rev.\ D {\bf 33}, 915 (1986).
  doi:10.1103/PhysRevD.33.915
  
  
  \bibitem{Parikh:1997ma}
M.~Parikh and F.~Wilczek, ``{An Action for black hole membranes},'' {\em Phys.
  Rev.}, vol.~D58, p.~064011, 1998.
  
  
   
  
\bibitem{Zhao:2015inu} 
  T.~Y.~Zhao and T.~Wang,
  ``Membrane paradigm of black holes in Chern-Simons modified gravity,''
  JCAP {\bf 1606}, no. 06, 019 (2016)
  doi:10.1088/1475-7516/2016/06/019
  [arXiv:1512.01919 [gr-qc]].
  
  \bibitem{groot1980relativistic}
S.~Groot, W.~Leeuwen, and C.~van Weert, {\em Relativistic kinetic theory:
  principles and applications}.
\newblock North-Holland Pub. Co., 1980.
  
\bibitem{Israel:1979wp} 
  W.~Israel and J.~M.~Stewart,
  ``Transient relativistic thermodynamics and kinetic theory,''
  Annals Phys.\  {\bf 118}, 341 (1979).
  doi:10.1016/0003-4916(79)90130-1
  
  
  \bibitem{cercignani2002relativistic}
C.~Cercignani and G.~Kremer, {\em The Relativistic Boltzmann Equation: Theory
  and Applications}.
\newblock Progress in Mathematical Physics, Birkh{\"a}user Basel, 2002.

\bibitem{wald1994quantum}
R.~M. Wald, {\em Quantum field theory in curved spacetime and black hole
  thermodynamics}.
\newblock University of Chicago Press, 1994.
  
\bibitem{Jacobson:2003wv} 
  T.~Jacobson and R.~Parentani,
  ``Horizon entropy,''
  Found.\ Phys.\  {\bf 33}, 323 (2003)
  doi:10.1023/A:1023785123428
  [gr-qc/0302099].


\bibitem{Rogatko:2002eu}
M.~Rogatko, ``{Physical process version of the first law of thermodynamics for
  black holes in Einstein-Maxwell axion dilaton gravity},'' {\em Class. Quant.
  Grav.}, vol.~19, pp.~3821--3827, 2002.
  

\bibitem{Amsel:2007mh}
A.~J. Amsel, D.~Marolf, and A.~Virmani, ``{The Physical Process First Law for
  Bifurcate Killing Horizons},'' {\em Phys. Rev.}, vol.~D77, p.~024011, 2008.



\bibitem{Bhattacharjee:2014eea} 
  S.~Bhattacharjee and S.~Sarkar,
  ``Physical process first law and caustic avoidance for Rindler horizons,''
  Phys.\ Rev.\ D {\bf 91}, no. 2, 024024 (2015)
  doi:10.1103/PhysRevD.91.024024
  [arXiv:1412.1287 [gr-qc]].
 
 \bibitem{Chatterjee:2011wj}
A.~Chatterjee and S.~Sarkar, ``{Physical process first law and increase of
  horizon entropy for black holes in Einstein-Gauss-Bonnet gravity},'' {\em
  Phys. Rev. Lett.}, vol.~108, p.~091301, 2012.

\bibitem{Kolekar:2012tq}
S.~Kolekar, T.~Padmanabhan, and S.~Sarkar, ``{Entropy Increase during Physical
  Processes for Black Holes in Lanczos-Lovelock Gravity},'' {\em Phys. Rev.},
  vol.~D86, p.~021501, 2012.

\bibitem{Bhattacharjee:2015yaa} 
  S.~Bhattacharjee, S.~Sarkar and A.~C.~Wall,
  ``Holographic entropy increases in quadratic curvature gravity,''
  Phys.\ Rev.\ D {\bf 92}, no. 6, 064006 (2015)
  doi:10.1103/PhysRevD.92.064006
  [arXiv:1504.04706 [gr-qc]].
  
\bibitem{Bhattacharjee:2015qaa} 
  S.~Bhattacharjee, A.~Bhattacharyya, S.~Sarkar and A.~Sinha,
  ``Entropy functionals and c-theorems from the second law,''
  Phys.\ Rev.\ D {\bf 93}, no. 10, 104045 (2016)
  doi:10.1103/PhysRevD.93.104045
  [arXiv:1508.01658 [hep-th]].

\bibitem{Jacobson:2011dz} 
  T.~Jacobson, A.~Mohd and S.~Sarkar,
  ``Membrane paradigm for Einstein-Gauss-Bonnet gravity,''
  Phys.\ Rev.\ D {\bf 95}, no. 6, 064036 (2017)
  doi:10.1103/PhysRevD.95.064036
  [arXiv:1107.1260 [gr-qc]].
  
\bibitem{Kolekar:2011gg} 
  S.~Kolekar and D.~Kothawala,
  ``Membrane Paradigm and Horizon Thermodynamics in Lanczos-Lovelock gravity,''
  JHEP {\bf 1202}, 006 (2012)
  doi:10.1007/JHEP02(2012)006
  [arXiv:1111.1242 [gr-qc]].
  
  
  
\bibitem{Chakraborty:2017kob} 
  S.~Chakraborty, A.~Ghosh and S.~Sarkar,
  ``Physical process first law for dynamical black holes and the membrane paradigm,''
  arXiv:1709.08925 [gr-qc].
  
  
  
\bibitem{Jacobson:1995uq} 
  T.~Jacobson, G.~Kang and R.~C.~Myers,
  ``Increase of black hole entropy in higher curvature gravity,''
  Phys.\ Rev.\ D {\bf 52}, 3518 (1995)
  doi:10.1103/PhysRevD.52.3518
  [gr-qc/9503020].
  
\bibitem{Wald:1993nt} 
  R.~M.~Wald,
  ``Black hole entropy is the Noether charge,''
  Phys.\ Rev.\ D {\bf 48}, no. 8, R3427 (1993)
  doi:10.1103/PhysRevD.48.R3427
  [gr-qc/9307038].

\bibitem{Iyer:1994ys} 
  V.~Iyer and R.~M.~Wald,
  ``Some properties of Noether charge and a proposal for dynamical black hole entropy,''
  Phys.\ Rev.\ D {\bf 50}, 846 (1994)
  doi:10.1103/PhysRevD.50.846
  [gr-qc/9403028].
  
\bibitem{Burger:2018hpz} 
  D.~J.~Burger, N.~Moynihan, S.~Das, S.~Shajidul Haque and B.~Underwood,
  ``Towards the Raychaudhuri Equation Beyond General Relativity,''
  arXiv:1802.09499 [gr-qc].
  
\bibitem{Brigante:2007nu} 
  M.~Brigante, H.~Liu, R.~C.~Myers, S.~Shenker and S.~Yaida,
  ``Viscosity Bound Violation in Higher Derivative Gravity,''
  Phys.\ Rev.\ D {\bf 77}, 126006 (2008)
  doi:10.1103/PhysRevD.77.126006
  [arXiv:0712.0805 [hep-th]].
  
\bibitem{Kovtun:2004de} 
  P.~Kovtun, D.~T.~Son and A.~O.~Starinets,
  ``Viscosity in strongly interacting quantum field theories from black hole physics,''
  Phys.\ Rev.\ Lett.\  {\bf 94}, 111601 (2005)
  doi:10.1103/PhysRevLett.94.111601
  [hep-th/0405231].
  
\bibitem{Camanho:2009vw} 
  X.~O.~Camanho and J.~D.~Edelstein,
  ``Causality constraints in AdS/CFT from conformal collider physics and Gauss-Bonnet gravity,''
  JHEP {\bf 1004}, 007 (2010)
  doi:10.1007/JHEP04(2010)007
  [arXiv:0911.3160 [hep-th]].
  
\bibitem{Buchel:2009sk} 
  A.~Buchel, J.~Escobedo, R.~C.~Myers, M.~F.~Paulos, A.~Sinha and M.~Smolkin,
  ``Holographic GB gravity in arbitrary dimensions,''
  JHEP {\bf 1003}, 111 (2010)
  doi:10.1007/JHEP03(2010)111
  [arXiv:0911.4257 [hep-th]].
  
\bibitem{Sarkar:2013swa} 
  S.~Sarkar and A.~C.~Wall,
  ``Generalized second law at linear order for actions that are functions of Lovelock densities,''
  Phys.\ Rev.\ D {\bf 88}, 044017 (2013)
  doi:10.1103/PhysRevD.88.044017
  [arXiv:1306.1623 [gr-qc]].

  
 
  
  \bibitem{PhysRevE.53.5808}
D.~J. Evans and D.~J. Searles, ``Causality, response theory, and the second law
  of thermodynamics,'' {\em Phys. Rev. E}, vol.~53, pp.~5808--5815, Jun 1996.
  

  
\bibitem{Camanho:2014apa} 
  X.~O.~Camanho, J.~D.~Edelstein, J.~Maldacena and A.~Zhiboedov,
  ``Causality Constraints on Corrections to the Graviton Three-Point Coupling,''
  JHEP {\bf 1602}, 020 (2016)
  doi:10.1007/JHEP02(2016)020
  [arXiv:1407.5597 [hep-th]].

\end{thebibliography}
\end{document}